\documentclass[preprint,12pt]{elsarticle}
\usepackage{graphicx}
\usepackage{subfigure}

\newcommand{\figwidth}{3.2 in}

\usepackage{epsfig}
\usepackage{float}
\usepackage[hidelinks]{hyperref}
\journal{Physics Letters B}
\begin{document}
\begin{frontmatter}
\title{On the origin of matter-antimatter asymmetry in the Universe}
\author{Efstratios Manousakis}
\affiliation{organization={Department of Physics, Florida State University and \\Department of Physics, National and Kapodistrian University of Athens, Panepistimioupolis, Zografos, 157 84 Athens, Greece},
  city={Tallahassee},
  post={32306-4350},
  state={Florida},
  country={USA}}
  
\begin{abstract}
  In order to investigate the origin of matter-antimatter
  asymmetry in   the Universe, we adopt a
  theoretical framework where the standard model emerges as a Poincar\'e invariant
   field theory localized at a domain-antidomain wall (DW-aDW)
  brane pair  of    a theory which lives on a
   higher dimensional bulk. We argue that such a system of a parallel DW-aDW pair could have been created at a very
   early epoch of the cosmological evolution when the Universe was still of microscopic size
   because its creation is topologically possible as compared to a single DW creation when the perpendicular extra dimension is compact. 
   The conservation laws, such as of charge and chirality, 
   are not violated in vacuum fluctuations of the combined DW-aDW system,
   but as we show their simultaneous conservation for each wall separately may not  be favorable in certain processes. In particular, as expansion of
   spacetime occurs in the higher dimensional bulk,
    the distance $d$ between the DW and the aDW increases as a
    function of time.
   We show that during the early stages of the cosmological evolution, when $d$ was of microscopic size,
   the leading mechanism  for pair-creation 
   from fluctuations of the gauge-field with polarization
   perpendicular to the DWs (i.e., polarization along the extra dimension)
   was one in which the  particle and the antiparticle were created on the opposite domain-walls hosting opposite chirality fermions.
   This mechanism allows for a matter-antimatter asymmetry to appear separately in the DW and the aDW, while in the combined DW-aDW system no such asymmetry was allowed.
     In this scenario, at a later and the present stage of the cosmological evolution, where $d$ is macroscopically large, the probability to
     violate these conservation laws on a single DW,  as a function of the inter-wall distance, is exponentially  suppressed.

\end{abstract}

\end{frontmatter}

\newpage
\section{Introduction}
The idea of using additional dimensions dates back to the early works of
Kaluza\cite{1921SPAW.......966K} and Klein\cite{1926Natur.118..516K}, who tried to unify electromagnetism with 
gravity by means of a theory with a compact fifth dimension. During the last two decades, the main attempts to use extra dimensions
aim at incorporating gravity and gauge interactions in a unique
scheme in a reliable manner. Extra dimensions is a fundamental aspect of
string theory\cite{GREEN1984117,doi:10.1142/S0217732389002331,PhysRevLett.75.4724} which is consistently formulated only in
a space-time of more than four dimensions. For some
time, however, it was conventional to assume that such extra dimensions were compactified to
manifolds of small radii, with sizes about the order of the Planck length, $l_P \sim 10^{-33}$ cm,
such that they would remain hidden to the experiment, thus, justifying why we see only four
dimensions\cite{strings}. In this picture, it was believed that the relevant energy scale where quantum
gravity and string effects would become important is Planck's mass $M_Pc^2$ from
where Planck's length is defined as $l_P =\hbar /(M_P c)$.
Later studies of the E$_8\times$E$_8$ heterotic string by Witten\cite{WITTEN1996135}  and Horava and Witten\cite{HORAVA1996506,HORAVA199694}
includes a systematic analysis of eleven-dimensional supergravity on a manifold with boundary believed to be relevant to this heterotic string.
These studies suggest that some, if not all, of the extra dimensions could be larger than the Planck length scale.
String theory constructions include D-branes in which, while
gravity can be free to propagate in extra dimensions, the standard model (SM)
fields are confined to branes.
For example Arkani-Hamed, Dimopoulos and
Dvali\cite{ARKANIHAMED1998263} (ADD) have introduced large extra dimensions on the order of
$mm$ scale in order to solve the
gauge hierarchy problem, although experiments probing the short range
part of Newton law of gravity have set limits of the order of $40 \mu m$ on
the size of the extra dimensions.
Randall and Sundrum (RS)\cite{PhysRevLett.83.3370} have given an alternative to compactification in string theory in a form of a warped extra dimension
confined between two branes. In this model, gravity is localized to one of the
boundaries and the deviations from Newtonian gravity become relevant
at very small length scales.

 There are 
  theoretical frameworks where the standard model emerges as a Poincar\'e invariant
  low-energy field theory localized at a 3+1 brane or a general defect of
   a theory which lives on a
   higher dimensional bulk. The localization of the various fields may happen
   by their individual coupling to different background fields that form
   these domains or branes.
   As was pointed out early\cite{PhysRevD.13.3398} DW branes select to bind fermions of specific chirality, and this led
   to many attempts to study the SM fields confined on a domain-wall brane, including those in Ref.~\cite{PhysRevD.77.124038} and in Refs.~\cite{doi:10.1142/S0217732319500809}
   \cite{PhysRevD.75.105007}. These investigations have been able to find mechanisms to localize fermions, scalar fields, and gauge fields on a
   DW produced when the vacuum expectation value of a background field is non-zero.

   In the present paper, we consider  a combination of a 3+1D DW-aDW pair
   that, we propose, emerged at the Big Bang in an expanding
   higher-dimensional spacetime (4+1D). We show that at a very early epoch of
   this evolution, when these DWs were sufficiently close as compared
   to the fermion DW localization length, matter and antimatter particles
   were created due to vacuum and thermal fluctuations at opposite
   DWs due to conservation laws. We use this finding as a scenario
   to explain the matter-antimatter asymmetry in our Universe.
   In the following Section we introduce this DW-aDW system and
   we show that it can emerge under the non-equilibrium conditions
   at the Big-Bang, and we illustrate how massless fermions can be hosted.
   In Sec.~\ref{baryon-asymmetry} we discuss the origin of the baryon-asymmetry, in Sec.~\ref{issues} we list remaining critical issues that need
   future investigation, and we present our conclusions in Sec.~\ref{conclusions}.
   
\section{Description of the DW-aDW system}

Our work is based on an action in a 4+1 Minkowski space with metric $g_{MN}$=
diag(1,-1,-1,-1,-1) where the Latin letters $MN$ are indices for the full
five dimensions and the Greek letters are indices for the 3+1 dimensional
subspace. The coordinates $x^{M}=(x^{\mu},x^4)$, where $x^0,x^1,x^2,x^3$
are the coordinates of the 3+1 dimensional spacetime and $x^4$ is the extra
dimension. The action in this five dimensional space is
\begin{eqnarray}
  {\cal S} &=& {\cal S}_{\eta} + {\cal S}_{\Psi} + {\cal S}_{I}, \\
  {\cal S}_{\eta} &=& \int d^5 x  \Bigl ({1 \over 2} {\partial}^M \eta \partial_M \eta - V(\eta)\Bigr ),\hskip 0.2 in 
V(\eta)= {1 \over 2} m_0^2 \eta^2 + {{\lambda} \over 4} \eta^4, \\
  {\cal S}_{\psi} &=& i \int d^5 x {\bar \Psi}  \Gamma^M \partial_M \Psi, \hskip 0.2 in 
  {\cal S}_{I} = - h \int d^5 x   \eta {\bar \Psi} \Psi,
\end{eqnarray}
We note that in
4+1D spacetime, the Clifford algebra consists of five Gamma matrices, with
the first four being the same ones as for 3+1D, i.e.,  $\Gamma^{\mu} = \gamma^{\mu}$ for $\mu=0,1,2,3$
and the fifth gamma matrix being $\Gamma^5= i \gamma^5$ where $\gamma^5$  is the 3+1D chirality operator.
Here $\eta$ in the scalar field with $\lambda>0$ and in its disordered phase
  $m_0^2\ge0$ and in its broken phase $m_0^2<0$. Therefore, we begin with massless fermions in 4+1 dimensions coupled
to a background scalar field $\eta$, which generates the domain walls, via the Yukawa coupling $-h {\bar \Psi} \Psi \eta$.

\begin{figure}[htp]
  \centering
  \includegraphics[width=\figwidth]{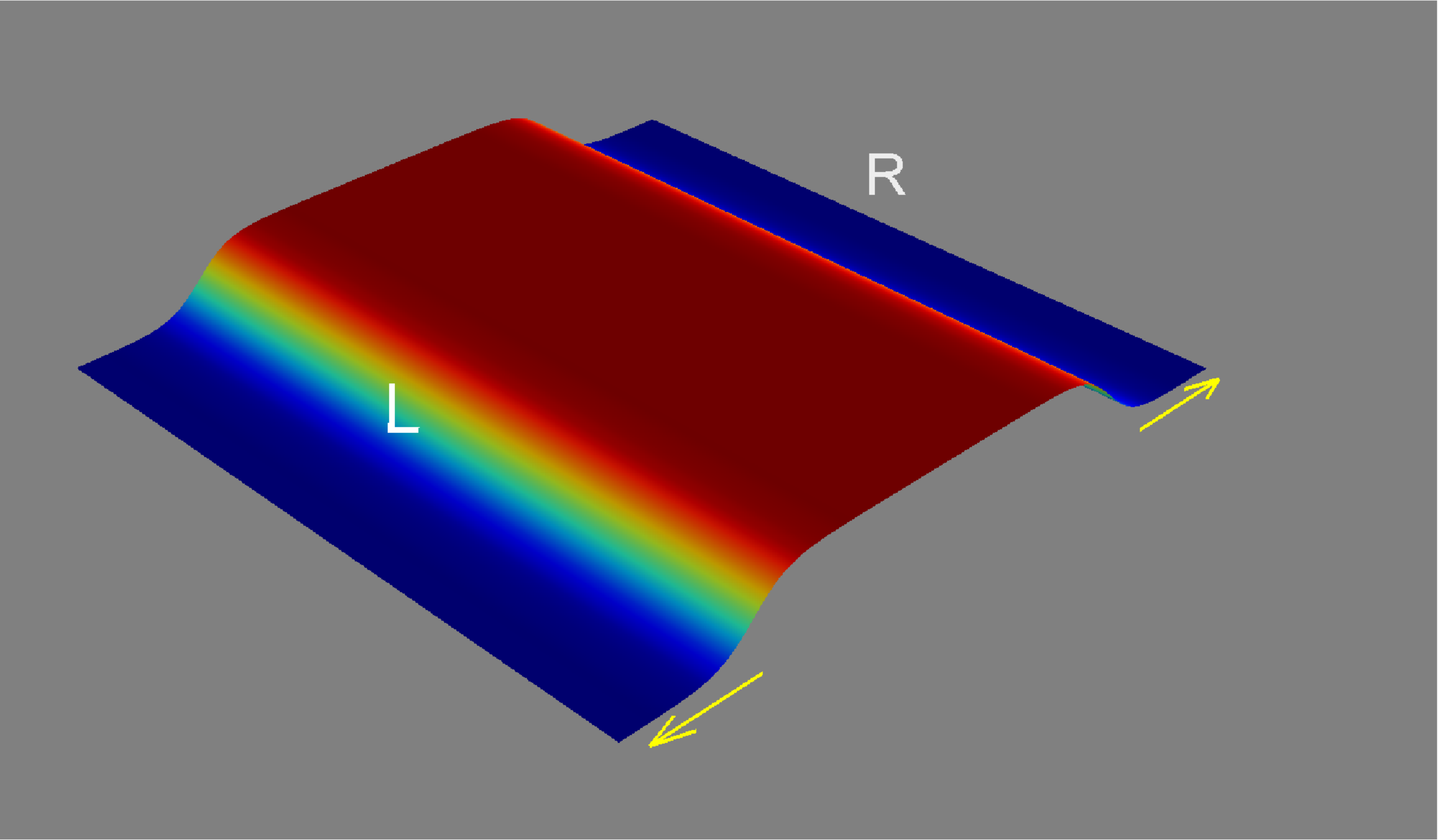}
  \caption{
    A pair of domain/anti-phase domain wall branes, created near each other at the GUT time scale,
    separated from one another as the cosmological evolution progressed.
    }
  \label{fig:domain-walls}
\end{figure}

It is well-known that if we seek the extrema of the action ${\cal S}_{\eta}$ in the broken phase, we encounter the equation 
\begin{eqnarray}
  -\eta^{\prime\prime}(x_4) + m_0^2 \eta + \lambda \eta^3 = 0,
\end{eqnarray}
where $\eta^{\prime\prime}(x_4)$ denotes the second derivative to $\eta(x_4)$ with respect to the coordinate $x_4$. In the broken phase where a uniform solution for the vacuum expectation value (VEV) of $\eta$ is given by
$v^2 = |m_0^2/\lambda|$,
we  also find non-uniform domain-wall like solutions $\eta(x_4)$ i.e., of the form
\begin{eqnarray}
  \eta(x_4) &=& v \tanh(\kappa (x_4-c)), 
 \label{eta_0}
\end{eqnarray}
where the $c$ denotes the location of the domain wall center in the $5^{th}$ dimension and the parameter $\kappa$ is
related to $v$ and the coefficient $\lambda$ of the $\phi^4$
term in the scalar field action as $\kappa = v \sqrt{\lambda/2}$.
It is also clear
that this equation can have as solutions a combination of a DW and of an aDW
as illustrated in Fig.~\ref{fig:domain-walls} if these are well separated. 

The initial state of the Universe was at a very high
temperature and, thus, it was not in the broken phase of the scalar
field.
Simulations\cite{PhysRevB.43.2615,PhysRevB.40.2205} of the $O(N)$ non-linear $\sigma$-model have
revealed the phenomena described below.
Let us assume that our simulation starts
with the system at a high temperature, well above that
corresponding to the broken phase, and subsequently the system is rapidly
quenched at a temperature below the transition temperature.
The initial disordered state of the system relaxes to its ordered
broken state in two stages and there are at least two very
different time-scales associated with such a two-stage relaxation
process.
First, different regions of the systems quickly freeze to
ordered configurations, which correspond to the
broken phase, but different regions choose different orientations
of the order parameter, which is fine because the free-energy is
$O(N)$ invariant. Namely, because the short-range interactions
are stronger, the system 
minimizes the contribution to the free-energy from these
interactions quickly, without paying much attention
to find a common global direction of the order parameter. This type
of information would take a longer time-scale to be communicated throughout
the various regions of the system.
An example is shown in Fig.~6 of Ref.\cite{PhysRevB.43.2615} for the
$O(3)$ non-linear $\sigma$-model in 2+1 dimensions.
Because of the $O(N)$ symmetry, 
which would be spontaneously broken
below the transition temperature, there is a
manifold of possible order-parameter orientations all
of which share the same magnitude that yields the minimum
action.
 The system attempts to align these domains
in the second stage of freezing, however, now this takes 
a much-much longer time-scale to reach an
equilibrium where these defects are mutually annihilated.
This state of randomly oriented ordered domains
corresponds to a very long-lived matastable state because the
time-scale grows exponentially with the domain volume.

This is also well-known in experimental physics 
when, in the absence of
an external field, we cool down an $O(N)$-invariant magnetic system, i.e., with no internal
anisotropy which can impose a preferred direction. Cooling such a system from its high-temperature
non-magnetic phase down to
the magnetically ordered phase,
the so-called Curie-Weiss magnetic domains form below the
critical temperature.

We expect the formation of these domain to also occur in higher dimensions.
In fact, we expect that these domains, which are mean-field solutions, should be more protected against
thermal and quantum fluctuations in higher dimensions.
In fact,  renormalization group calculations based on  the
$\epsilon=4-d$ expansion\cite{wilson} show that above the upper critical dimension $d_c=4$ the mean-field like picture is expected to hold.

Therefore, one would expect a similar scenario to happen
in the initially very hot Universe which undergoes rapid cooling.
We would expect several DW-aDW pairs, i.e., twists of
the local scalar-field VEV (such as those
illustrated in Fig.~\ref{fig:domain-walls}), to appear  as the temperature
cools below the critical temperature which corresponds
to its broken phase. We would assume
that our Universe exists as one DW, which appears next to an aDW as schematically illustrated in Fig.~\ref{fig:domain-walls}.
In the present work, we have assumed a compact geometry, i.e., periodic boundary
conditions (PBC) along the fifth dimension.
   The total phase of the VEV of the background field, which behaves as an order parameter, changes by $\pi$ when crossing a single domain
   wall and this gives a non-zero value in a type of topological charge
   associated with the order-parameter phase. However, the total topological charge when we have a combination of
   a DW-aDW pair is zero and thus, this can occur as a
   non-equilibrium spontaneous vacuum fluctuation of the background field at an early epoch of the cosmological evolution
   when the Universe was hot and rapidly cools below a characteristic
   temperature and enters the ordered phase (like two Weiss domains).
   On the other hand, just a single DW can not emerge spontaneously
   in the case where the extra dimensions are compactified. In this
   case,   DWs can only occur in DW-aDW pairs, where the VEV to right of
   the right DW (blue domain of Fig.~\ref{fig:domain-walls}) and to the left of the left aDW (also blue domain of Fig.~\ref{fig:domain-walls}) are the same and the can
   match when PBC (compact) are  applied.  We will assume that the two domain walls
   were naturally close at the time of their emergence and
   the distance $d$ between them increases  during the cosmological expansion
   in the same way as the 3+1 DW expands. 
Namely, it is natural to assume that the ``fabric'' of
of the 3+1D DW expands, as observed, because the 4+1 D bulk expands as a whole.

 This leads to the following Dirac
equation in 4+1 dimensions:
\begin{eqnarray}
  ( i \gamma^{\mu} \partial_{\mu} + \gamma^5 \partial_4 - h \eta) \Psi =0,
  \label{eq:Dirac}
\end{eqnarray}
where $x_{\mu}$, $\mu=0,1,2,3$ are the four components of the
3+1 spacetime coordinates and $x_4$ is the extra dimension.
Let us choose the origin of the extra dimension to be in the
middle between the DW and the aDW so that the locations of
the left DW $x_L$ and of the right aDW $x_R$ respectively are
related as $x_L=-x_R$ and $x_R>0$.
We will express the  background scalar field $\eta$ which couples
to the fermions  as
\begin{eqnarray}
  \eta(x_4<0) &=& v \tanh(\kappa (x_4-x_L)), \nonumber \\
  \eta(x_4>0) &=& -v \tanh(\kappa (x_4-x_R)), \label{eta}
\end{eqnarray}
The above Eq.~\ref{eta} is a combination of an anti-phase
domain walls separated by a distance $d=|x_R-x_L|$.   When
$d\to \infty$ the two domain walls are each a saddle point solution
to the scalar field action. Our function $\eta$ is schematically illustrated in
Fig.~\ref{fig:domain-walls} for the 2+1 dimensional case where
visualization is possible.
In the present work, when we discuss the
cosmological evolution, the Universe, as a whole at early epochs, is considered bound as it expands. Therefore, 
Fig.~\ref{fig:domain-walls} should be considered as a drawing of a
portion of the DW-aDW, i.e., from a local viewpoint, and it
should not imply that the 4D bulk is infinitely extended.

We can imagine a scenario where these two
domain walls were created very close at the  early stages of the
cosmological evolution (because out of nothing only pairs of domain walls
of opposite topological charge can be created) and they are separated
from each other at a later time. We want to use a framework to
understand this evolution, so it is convenient to use as basis
the state  of the well-separated domain walls. 
The reason for considering a pair of such anti-phase domain walls is that
each domain wall is characterized by a non-zero topological
charge associated with the background field configuration.
The pair has zero topological charge and their creation is
easier to imagine, whereas creating just an isolated
domain wall requires changing the resting configuration from
$-\infty$ all the way to $+\infty$. The pair of opposite topological
charges belongs to the same topological sector as the flat configuration,
whereas the single domain wall configuration is separated
from the no-domain wall configuration by an infinite action barrier.

We will assume that the two domain walls
were naturally close  to each other at the time of their creation and they
separated as the time went on because of the
expansion in the same way as the 3+1 DW expands. We would like to analyze the early stage of the
domain/anti-domain wall system in terms of the easier to
conceptualize asymptotic states of two infinitely
separated domain-walls. This approach is
similar to that in solid state or in molecular physics where the
state of many atoms is analyzed in terms of hopping of electrons
between atomic orbitals, the so-called tight-binding approximation\cite{Manousakisbook}.
In the case where the DW is far away from aDW, we write the solution of the Dirac
Eq.~\ref{eq:Dirac} as
\begin{eqnarray}
  \Psi(x_{\mu},x_4) = \psi(x_{\mu}) f(x_4),
\end{eqnarray}
where the spinor $\psi(x_{\mu})$ describes a 3+1 dimensional
massless fermion, $i \gamma^{\mu} \partial_{\mu} \psi = 0$,
which is localized on either of the domain walls, and the function
$f(x_4)$ specifies where the fermion is localized along the
extra dimension $x_4$ and it is obtained as a solution to the following
differential equation:
\begin{eqnarray}
  (  f^{\prime} \gamma^5 - h \eta f) \psi =0.
  \label{eq:dw1}
\end{eqnarray}
We can choose our basis for the spinor part $\psi$ to be the eigenstates of the chirality operator $\gamma^5$, for
massless fermions, i.e.,
$\gamma^5 \psi^{+} =  \psi^{+}$, and $\gamma^5 \psi^{-} = - \psi^{-}$, and
in this case the above equation splits into the following two:
\begin{eqnarray}
  f^{-\prime}  + h \eta f^-  &=&0, \label{eq:dwa}\\
  f^{+\prime} - h \eta f^+  &=&0,
  \label{eq:dwb}
\end{eqnarray}
and the general solution to Eq.~\ref{eq:dw1} may be approximately
written as follows:
\begin{eqnarray}
  \Psi = f^-(x_4) \psi^-(x_{\mu}) + f^+(x_4) \psi^+(x_{\mu}),
\end{eqnarray}
where $f^-$ and $f^+$ are the most general solutions to the
Eqs.~\ref{eq:dwa},\ref{eq:dwb}. This approximation makes sense
when the two DWs are well-separated, i.e., when $\kappa |x_R-x_L|>>1$.

We divide the extra dimension in two regions , the region
of $x_4$ near the left domain wall, i.e., region $L$ ($x_4<0$) and
the case where $x_4>0$ (region $R$).
We can integrate Eqs.~\ref{eq:dwa},\ref{eq:dwb} and find that
in region $L$ ($x_4 <0$) 
\begin{eqnarray}
  f_{L}^-(x_4<0) &=& {c^-_L} \xi(x_4-x_L), \\
  f_{L}^+(x_4<0) &=& {{c^+_L} \over { \xi(x_4-x_L)}},\\
\xi(x) &\equiv&  [\cosh(\kappa x)]^{-h v /\kappa}.
\end{eqnarray}
However, because of the wavefunction normalizability condition, we need to
choose $c^+_L=0$.
Similarly for region $R$, the solutions are of the form
\begin{eqnarray}
  f_{R}^-(x_4>0) &=& {{c^-_R} \over { \xi(x_4-x_R)}}, \\
  f_{R}^+(x_4>0) &=& {c^+_R} {\xi(x_4-x_R)}.
\end{eqnarray}
The normalizability condition in this case yields $c^-_R=0$.
Therefore, there are only left-handed acceptable solutions in the left
domain wall,  i.e.,
\begin{eqnarray}
  { U}^-_{L} &=& {C^-_{L}} \xi(x_4-x_{L}) e^{-i(E_+x_0-{\bf p}\cdot {\bf x})} u^-_{L},\label{eq:ul}\\
  { V}^-_{L} &=& {D^-_{L}} \xi(x_4-x_{L}) e^{-i(E_-x_0-{\bf p}\cdot {\bf x})} v^-_{L},\label{eq:vl}
\end{eqnarray}
  and right-handed acceptable solutions in the right domain wall,
\begin{eqnarray}
  U^+_{R} &=& {C^+_{L}} \xi(x_4-x_{R}) e^{-i(E_+x_0-{\bf p}\cdot {\bf x})} u^+_{R},\label{eq:ur}\\
  V^+_{R} &=& {D^+_{L}}  {\xi(x_4-x_{R})} e^{-i(E_-x_0-{\bf p}\cdot {\bf x})} v^+_{R},\label{eq:vr}
\end{eqnarray}
where $u^s_{L}$ ($u^s_{R}$) are chirality-$s$ positive-energy solutions  localized on the left (right) DW
and $v^s_{L}$ ($v^s_{R}$) are chirality-$s$ negative-energy solutions localized on the left (right) DW. The constants  $C^{\pm}_{L,R}$ and $D^{\pm}_{L,R}$
are coefficients to be determined by the wavefunction normalization condition.
When the two DWs 
are close to each other, the wavefunctions localized on the left and right DWs begin to overlap. In this case,
these wavefunctions, which are localized at the
centers $x_R$ and $x_L$, can be used as an unperturbed basis to define
an effective $4 \times 4$ Hamiltonian where the other degrees of freedom from the extra dimension are implicitly integrated out.
First,  the zero point motion of the scalar field around this fixed field configuration has been integrated out. Namely, small amplitude 
fluctuations of this type can be also included, by carrying out path integration
of Gaussian fluctuations around the extrema (the DW-aDW solutions).
These will only change the matrix
elements $V$ and $U$ by a prefactor. However,  our qualitative analysis does not
rely on specific values of these amplitudes.

The role of higher-energy fermion excitations seem to have
also been excluded from the calculation. However, one can
imagine that this calculation is based on a quasi-degenerate
perturbation theory calculation\cite{PhysRevB.48.1028,Manousakisbook} where our model subspace is defined by
a projection operator $\hat M = |1 \rangle \langle 1 | +
|2 \rangle \langle 2 | + |3 \rangle \langle 4 | + |4 \rangle \langle 4 | $,
which 
is formed by these four states, and all the other states form a
space $\hat Q  = \sum_{n \ne 1,2,3,4} | n \rangle \langle n |$.
Then, the effective Hamiltonian, which operates only in the reduced
subspace spanned by these 4 states  is obtained
by including the virtual excitations to states outside this subspace
perturbatively. The inclusion of these states only modifies the
values of amplitudes $V$ and $U$. In the present paper,  we assumed that
by integrating out these virtual excitations we remain
in this reduced 4-dimensional subspace. Again, our qualitative analysis does not
depend on the specific values of the matrix elements $V$ and $U$ used above.

\begin{figure}[htb]
    \begin{center}
        \subfigure[]{
            \includegraphics[width=0.40\textwidth]{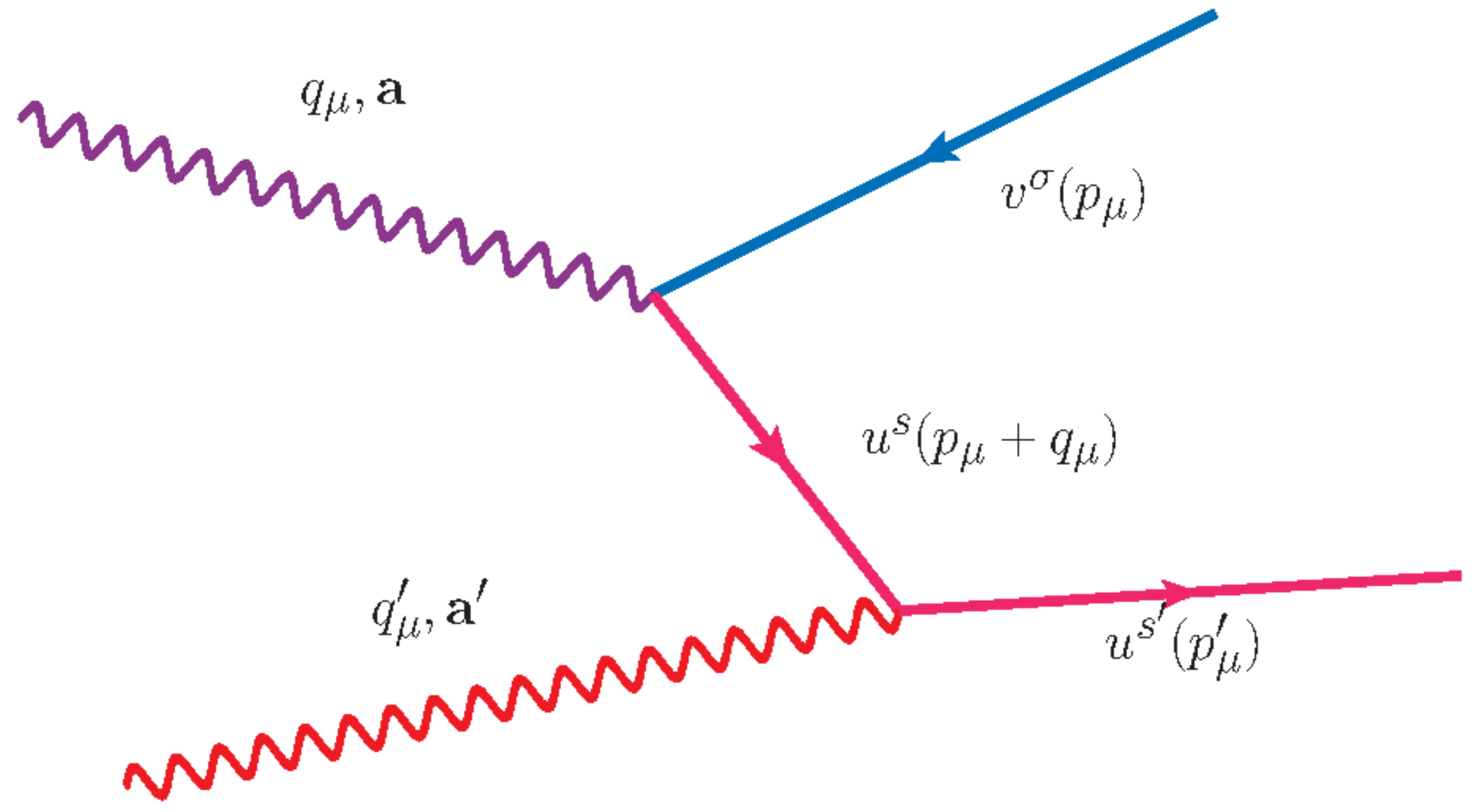}
          \label{fig:simple-pair}
        } \\
\vskip 0.2 in
        \subfigure[]{
            \includegraphics[width=0.40\textwidth]{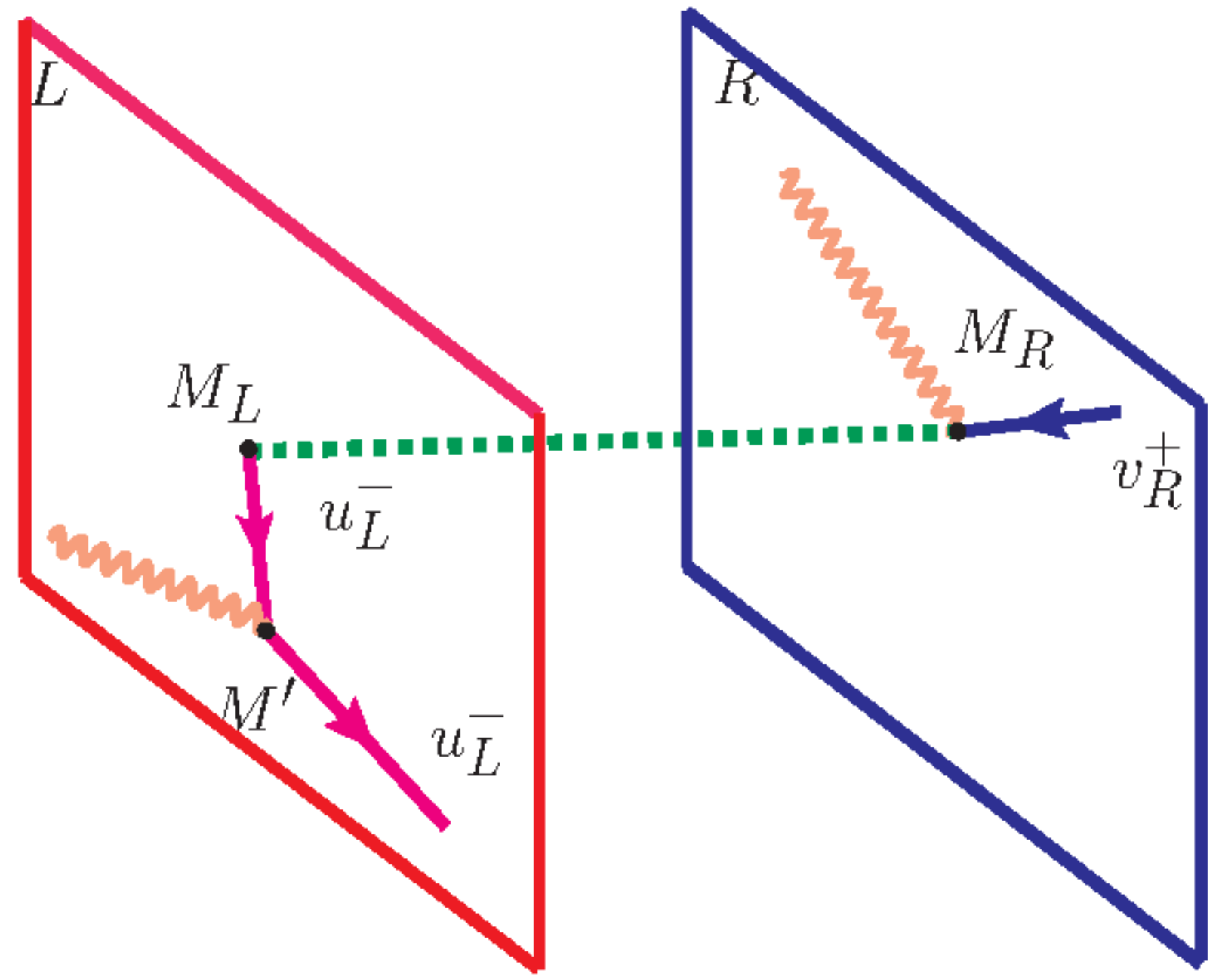}
          \label{fig:pair-process}
        } 
    \end{center}
    \caption{(a) The well-known pair production process. (b) The pair production where the antiparticle is created by
      a photon on the right DW and at the same time a particle is created on the left DW. The rest of the process
    on the left DW is similar to that in (a).}
        \label{fig:pair-production}
\end{figure}
\section{Generation of the baryon asymmetry}
\label{baryon-asymmetry}

Localizing the gauge field on the DW branes giving the correct SM on the brane might be tricky\cite{DVALI199764}.
Here, because of the requirement of gauge invariance
we will consider the minimal coupling of a gauge field to the fermions, i.e.,
\begin{eqnarray}
  \hat H_{int} &=& {\bf J} \cdot {\bf A},\\
  {\bf J} &=& e \Psi^{\dagger} \hat {\vec \alpha} \Psi,
\end{eqnarray}
and we are working in a gauge where $A_0=0$, and $\hat \alpha_i$, $i=1,2,3$ are the three Dirac matrices, and
\begin{eqnarray}
\hat \alpha_4 &=& -i \gamma^0 \gamma^5 = -i \left ( \begin{array}{cc} 
~~{\bf 0} & \hat {\bf 1}  \\
-\hat {\bf 1} & {\bf 0}\end{array} \right ).
\end{eqnarray}
This form of the current is consequence of Noether's theorem.
At the early epoch after the Big-Bang the high-energy photons are believed to create particle/antiparticle pairs. Let us consider
the process which generates electron-positron pairs as an example.
The first order process of a single photon creating an electron-positron pair cannot simultaneously conserve energy and momentum.
The leading process is a second-order process where two photons produce an electron-positron pair.
Following Dirac\cite{dirac_1930} we consider this process in 4+1 dimensions in the presence of the DW-aDW pair discussed previously.
We consider the field produced by two photons localized on one domain-wall and propagating on the 3+1 brane $x_{\mu}$. However, their
polarization vectors ${\bf a}_{1,2} $, which are perpendicular to their momenta ${\bf q}_{1,2}$ due to gauge invariance, can be
perpendicular to the domain-wall, i.e., along the dimension $x_4$. The vector potential can be written as
\begin{eqnarray}
   A_{\mu} &=& \Bigl [{\bf a}_{\mu L}^{\dagger} e^{-i (q_L x_0- {\bf q}_L \cdot {\bf x})} + {\bf a}_{\mu L} e^{i (q_L x_0- {\bf q}_L \cdot {\bf x})} \Bigr ] \zeta_{\mu L} \nonumber \\
      &+& \Bigl [{\bf a}_{\mu R}^{\dagger} e^{-i (q_R x_0- {\bf q}_R \cdot {\bf x})} + {\bf a}_{\mu R} e^{i (q_R x_0- {\bf q}_R \cdot {\bf x})} \Bigr ] \zeta_{\mu R},
          \end{eqnarray}
where ${\bf a}_L$ (${\bf a}_R$) is along  the polarization vector $\epsilon_{\mu}$ of the photon 
localized on the $L$ ($R$) domain-wall. There are three possible directions
of polarization, and they are all perpendicular to the
direction of momentum, which is assumed to lie on the 3+1 brane.
We will consider the role of the gauge field localized on the DWs,
but on one of the DWs, say the ${\bf a}_R$, 
having polarization along the extra dimension, i.e.,  that ${\bf a}_R$
is along $x_4$, while on the other DW, ${\bf a}_L$ can be with polarization
vector with components in the DW.
The function $\zeta_{L,R}=g(x_4-x_{L,R})$ is the profile of a
gauge-field localization function along the dimension $x_4$.
There are several matrix-elements of $\hat H_{int}$ between the states listed in Eqs.~\ref{eq:ul},\ref{eq:vl},\ref{eq:ur},\ref{eq:vr}.
Owing to the fact that
\begin{eqnarray}
  \hat \alpha_4  u^+ = -i  v^-, \hskip 0.2 in \hat \alpha_4 u^- = i v^+,\\ 
  \hat \alpha_4  v^+ = -i u^-, \hskip 0.2 in \hat \alpha_4 v^- = i u^+,
  \end{eqnarray}
tracing over the degrees of freedom of the $x_4$ direction, the following matrix elements of the 4th component of the current operator are non-zero:
\begin{eqnarray}
  \langle V^-_L|  J_4 | U^+_{R} \rangle, \hskip 0.2 in \langle V^+_R | J_4 | U^-_L \rangle,\\ 
  \langle U^-_L|  J_4 | V^+_{R} \rangle, \hskip 0.2 in \langle U^+_R |  J_4 | V^-_L \rangle,
  \end{eqnarray}
which are proportional to the overlap integral $I_f$ of the wavefunction profiles along the $x_4$ direction of the
fermion parts:
\begin{eqnarray}
I_f = C^{\pm}_R [C^{{\mp}}_L]^*&\int& {dx_4} \xi(x_4-x_L) \xi(x_4-x_R).
\end{eqnarray}
Assuming that the DWs are close enough such that this integral
has a significant value, the matrix elements above are non-zero.
In order to have a non-zero vertex for processes such as 
the one illustrated in Fig.~\ref{fig:pair-process}, i.e., for example,
to create an electron on the right DW and a positron on the
left DW due to an incident photon on the left  DW, we need
to consider the matrix element of the gauge field operator that contains
the term
\begin{eqnarray}
  \langle 0 | a^{\dagger}_{4,L} a^{\dagger}_{4,R} | 0 \rangle.
\end{eqnarray}
This matrix element is proportional to the following overlap integral
\begin{eqnarray}
  I_b =                \int dx_4 \zeta(x_4-x_L) \zeta(x_4-x_R).
  \end{eqnarray}
In order for the vertex to split over the two DWs (shown as $M_L$ and $M_R$
in Fig.~\ref{fig:pair-process}), the DW  needs to
be at a distance from the aDW less than the spread of the localization function of fermions 
$\xi(x)$ and of the gauge fields $\zeta(x)$.
The rest of the process shown in Fig.~\ref{fig:pair-process} which
takes place on the left DW in our example, is the same as in Fig.~\ref{fig:simple-pair}. This part requires the polarization of the ${\bf a}^{\dagger}_{R}$ to
be in the left DW brane dimensions.
Notice that the momentum, energy, charge and parity
conservation is fulfilled
if we consider all the inter-DW parts together. They 
are not conserved on each DW individually.
\section{Discussion of critical issues}
\label{issues}
As mentioned in the Introduction, there are several domain-wall models
that have been proposed by various investigators in order to explain
the emergence of the 3+1 dimensional space out of a higher dimensional
space.
The goal of the present paper was to build on the domain-wall models
aiming to obtain one that is more consistent with 
statistical physics
and thermodynamics and to explain the matter-antimatter asymmetry.
Embedding this braneworld scenario in a consistent UV theory and providing a full explanation of the cosmological evolution are challenging issues and
are beyond the scope of the present paper.
Such questions are not  unique to our model,
they are issues for all the domain-wall braneworld models and they
are currently being addressed by various investigations.
Below we list some of these critical issues of these models and of our present
proposal.

There are various different philosophies on how to
localize the gauge field on the DW brane.
Dvali and Shifman\cite{DVALI199764} relies on the non-perturbative nature of
confinement: they consider strongly-coupled 
gauged theories which can be in the non-Abelian confining phase outside the DW and in the Abelian Coulomb phase inside the DW. This
picture is complementary to  the dual superconductor model of confinement
first proposed by t'Hooft and Mandelstam where confinement
arises due to a magnetic monopole condensate\cite{PESKIN1978122}.
In  other attempts\cite{doi:10.1142/S0217732319500809,10.1143/PTP.124.71} 
a gauge coupling that depends on the
extra dimensions is used where, when trying to separate the
extra component of the gauge field from the other four components
$A_{\mu}$, it is found that the relation between
the $A_{\mu}$  and its fifth component in the 4-dimensional
effective theory is the one between a massive gauge boson and a would-be
Goldstone  mode after the spontaneous gauge symmetry breaking.
In this case, the KK-modes of the  fifth component are identified as the
Goldstone modes which would be eaten by the KK-modes of $A_{\mu}$.
This can happen if the parameters are chosen properly and after the
symmetry breaking, which happened at the GUT scale.

On the other hand, the scenario, which we describe in the present work,
is for the epoch before the GUT scale.
Another scenario might be that the fluctuations which we describe in this
work correspond to other KK modes and not to the zero mode.

Critical issues are the details of the expansion of this 4+1 dimensional
braneworld and  how the evolution of the
DW-aDW system couples to the inflation and the rest of the
cosmic evolution. This connection is very important, precisely because 
inflation dictates that
baryon asymmetry must be dynamically
generated after reheating, necessitating a mechanism of baryogenesis.
We have generalized the expansion of our Universe to an expanding
4+1 dimensional spacetime where the 3+1D DW (our Universe) and aDW
necessarily expand.
One important difference from other braneworld models
is that in the present scenario  we need to
consider the role of the characteristic time-scale when,
because of  this expansion,
the DW-aDW separation $d$ becomes of the order of or greater than
the DW thickness,
i.e., 1/$\kappa$.
The Planck data\cite{refId0} from the cosmic microwave radiation and
the current information on the elemental density in the Universe should be
analyzed to determine how this time-scale is related to the hierarchy of
the various other time-scales of the cosmological evolution, such as, the
inflation epoch as well as the inflaton density evolution.
Furthermore, new information from the detection of gravitational waves
imposes further constraints on the present and other braneworld
models\cite{PhysRevD.103.044031,PhysRevD.101.123537,PhysRevD.97.064039}. 
Therefore, an investigation of the relationship of the scenario presented in this paper where more special-case scenarios are discussed and
the inflation is urgently needed.

\section{Conclusions}
\label{conclusions}

The mechanism described here allows for a matter-antimatter
asymmetry to appear separately in the DW and the aDW, without
allowing such asymmetry to occur in the combined DW-aDW
system. Therefore, when the Universe was in a very high temperature state, where the occurrence of pair creation events should
be considered statistically independent, a matter-antimatter asymmetry is allowed to happen in each of the two DW subsystems.
Given that more than one type of SM particle-antiparticle creation
processes take place, the overall charge neutrality on each DW can
be maintained not necessarily by means of having equal number
of particles and anti-particles of each type. In addition, momentum,
energy, charge and parity conservation are all fulfilled if we
consider the DW-aDW system together.

The mechanism does not rely on the particular means
of the gauge field localization. It is only using the fermion field localization
on a DW-aDW pair.
We see no reason that by localizing the gauge field on a DW brane would
mean that the gauge field cannot have polarization along  the
extra dimension in the early epoch of the cosmological evolution.
However, as discussed earlier, this problem of localizing
the gauge field on the DW is tricky.

Photons with polarization along the extra dimension,
in order to be absorbed by a fermion
bound to the DW brane,  require a change in the chirality of the particle.
The only first order process of this type is the one
discussed in this work involving a photon with such polarization,
but it was only applicable when the Universe
was of microscopic size, i.e.,  when the matter localization length
on the DW was comparable to the DW-aDW separation.
This process is exponentially suppressed as  a function of the
DW-aDW distance and, thus,  in a Universe where the
DW and the aDW are well-separated, it is exponentially unlikely. 
As a result, when the DW and the aDW are well-separated,
these photons are not allowed to interact with
standard model particles except that they might contribute to the total
energy density  that causes  curvature.
  There are various scenarios to include
  their role in the
  cosmological evolution in order to explain
  the  high-resolution Planck data\cite{refId0}.
  These, along with addressing the critical issues listed in the
  previous Section, require a detailed future
  study.

  \section{Acknowledgments}
  The author would like to thank W. H. Green for an enlightening discussion about
  the analysis of the high resolution Planck data of the cosmic microwave background
  radiation.
\bibliographystyle{elsarticle-num-names}

\end{document}